\documentclass[12pt]{spieman}  % 12pt font required by SPIE;
\usepackage{amsmath,amsfonts,amssymb}
\usepackage{graphicx}
\usepackage{setspace}
\usepackage{tocloft}
\usepackage{xcolor}

\title{Sieging HELM's deep: PRIMA unveils the far-infrared properties of highly extincted low-mass galaxies}

\author[a,*]{Laura Bisigello}
\author[b]{Carlotta Gruppioni}
\author[c]{Giulia Rodighiero}
\author[c]{Giovanni Gandolfi}
\author[d]{James M. S. Donnellan}
\author[d]{Seb Oliver}
\author[d,e]{Stephen M.~Wilkins} %
\author[f]{{L. Y. Aaron} {Yung}}
\affil[a]{INAF--Osservatorio Astronomico di Padova, Vicolo dell'Osservatorio 5, I-35122, Padova, Italy}
\affil[b]{INAF-Osservatorio di Astrofisica e Scienza dello Spazio, via Gobetti 93/3, I-40129, Bologna, Italy}
\affil[c]{Dipartimento di Fisica e Astronomia "G. Galilei", Universit\`a di Padova, Via Marzolo 8, 35131 Padova, Italy}
\affil[d]{Astronomy Centre, University of Sussex, Falmer, Brighton BN1 9QH, UK}
\affil[e]{Institute of Space Sciences and Astronomy, University of Malta, Msida MSD 2080, Malta}
\affil[f]{Space Telescope Science Institute, 3700 San Martin Drive, Baltimore, MD 21218, USA}

\cftpagenumbersoff{figure}
\cftpagenumbersoff{table} 
\begin{document} 
\maketitle

\begin{abstract}
Although the majority of star-forming galaxies show a tight correlation between stellar mass and dust extinction, recent James Webb Space Telescope observations have revealed a peculiar population of Highly Extincted Low-Mass (HELM) galaxies, which could revolutionise our understanding of dust production mechanisms. To fully understand the dust content of these galaxies, which are a minority of the overall galaxy population, far-infrared observations over large areas are pivotal. In this paper, we derive the expected PRIMAger, the far-IR (24-235$\mu m$) imaging camera proposed for the Probe far-IR Mission for Astrophysics, fluxes for a set of photometric candidates HELM galaxies. Taking into account a deep survey of 1000\,h over $1\rm\, deg^{2}$, we expect to detect around $3.1 \times 10^4$ HELM sources in at least one PRIMAger filter, 100 of which are at $z=1-1.5$. For 32\% of this sample, there will be observations in at least four PRIMAger filters, covering at least the 90 to $240\rm\, \mu m$ wavelength range, which will allow us to obtain a detailed fit of the dust emission and estimate the dust mass.
\end{abstract}

% Include a list of up to six keywords after the abstract

% Include email contact information for corresponding author
{\noindent \footnotesize\textbf{*}  \linkable{laura.bisigello@inaf.it} }

\begin{spacing}{2}   % use double spacing for rest of manuscript

\section{Introduction}
\label{sect:intro}  % \label{} allows reference to this section
Even if dust is a minor component of the interstellar medium (ISM), accounting for 1\% of its total mass \cite{Draine2003}, it is crucial for the formation and cooling of the molecular gas and therefore on the process of star formation. Moreover, dust absorbs ultraviolet (UV) and optical light from newly formed stars, re-emitting it at infrared (IR) wavelengths \cite{Hauser2001}, impacting the estimation of physical properties. The impact of dust absorption is evident, for example, when looking at the cosmic star formation rate density, which is severely underestimated by 50\% at least up to $z=4$ if it is based only on UV data \cite{Madau2014,Gruppioni2020,Zavala2021,Traina2024}. In addition, an increasing number of works have identified sources faint or dark in the optical and bright at IR or radio wavelengths \cite{Wang2019,Gruppioni2010,Talia2021,Gentile2024}, due to their large dust obscuration, making IR observatories pivotal to have a complete view of the most obscured phases of galaxy evolution.

Although previous studies have shown the presence of a positive correlation between dust attenuation ($A_{V}$) and stellar mass for star-forming galaxies \cite{Pannella2015,McLure2018,Shapley2023,Liu2024}, implying that more massive star-forming galaxies are typically more dust-obscured; recent discoveries have challenged this understanding \cite{Rodighiero2023, Bisigello2023b,Bisigello2024,Gandolfi2025}. These studies, thanks to photometric and spectroscopic data from the James Webb Space Telescope (JWST), have revealed the intriguing population of Highly Extincted Low Mass (HELM) galaxies, characterised by stellar masses typically lower than $M^{*}<10^{8.5} M_{\odot}$, but a significant dust attenuation ($A_{V} > 1$). The existence of HELM galaxies contradicts the traditional understanding that low-mass galaxies should generally have a dust-poor, chemically unevolved ISM. Indeed, dust is thought to be primarily produced by mass loss from evolved stars and supernovae ejecta \cite{{Gall2011,Sarangi2018}}, indicating that an increase in star formation results in an increase in stellar and dust mass. Additionally, the lower gravitational potential of low-mass galaxies allows for efficient dust expulsion by stellar-driven outflows \cite{Dayal2013}. Therefore, the discovery of HELM galaxies open a question on how such low mass galaxies could produce and retrieve such large dust content and their study could therefore revolutionise our understanding of dust production mechanisms. 

The nature of HELM galaxies and the reason behind their large dust reservoir are still completely uncertain, given that only one of them has been confirmed spectroscopically \cite{Bisigello2024}, while photometric observations are limited to near-IR wavelengths with JWST. The dusty and low-mass nature of these HELM candidates has been derived fitting the available optical and near-IR photometric points and is therefore prone to degeneracies. Far-IR (FIR) observations are therefore essential to investigate the nature of these peculiar objects; they can directly measure the dust mass and temperature by detecting thermal emission from the dust, providing crucial insights into both dust properties and obscured star formation. Given that HELM galaxies correspond to a minority of the overall galaxy population (i.e. 4\%, Bisigello et al. in preparation), it is crucial to cover large areas with sensitive FIR observations to obtain a statistical sample of these objects. 

A pivotal mission to study HELM sources, given the wavelength coverage and the survey capability, is the Probe far-IR Mission for Astrophysics (PRIMA, Glenn et al., this volume). PRIMA is a FIR observatory concept of a telescope cryogenically-cooled at 4.5\,K and with a primary mirror with a diameter of 1.8\,m. Particular suited to study HELM sources over large areas is the imaging camera, called PRIMAger (Ciesla et al., this volume), which covers between 24 and 84$\,\rm \mu m$ with a hyperspectral imager, and between 96 and 235$\,\rm \mu m$ with four broad-band filters. The hyperspectral imager will have a field-of-view (FoV) of $3.8'\times3.0'$ and a full-width-half-maximum (FWHM) between $4''$ and $6''$, while the broad band filters will have a FoV of $5.0'\times4.5'$ and a FWHM ranging from $9''$ to $24''$. The large FoV will be crucial to observe wide sky areas. 

In this paper we make use of the HELM catalogue in the Cosmic Evolution Early Release Science Survey
(CEERS \cite{Bagley2023,Yang2023,Finkelstein2024}) to estimate the flux expected for the PRIMAger filters and the capability of this instrument to detect them. In particular, in Section \ref{sec:cat} we describe the input catalogue, while we present the prediction for PRIMAger in Section \ref{sec:predictions}. We summarize our results in Section \ref{sec:summary}. 
Throughout this paper, we consider a $\Lambda$CDM cosmology with $H_0=70\,{\rm km}\,{\rm s}^{-1}\,{\rm Mpc}^{-1} $, $\Omega_{\rm m}=0.27$, $\Omega_\Lambda=0.73$, a Chabrier initial mass function \cite{Chabrier2003}. 

\section{Input HELM catalogues}\label{sec:cat}
\begin{figure}
    \centering
    \includegraphics[width=0.75\linewidth]{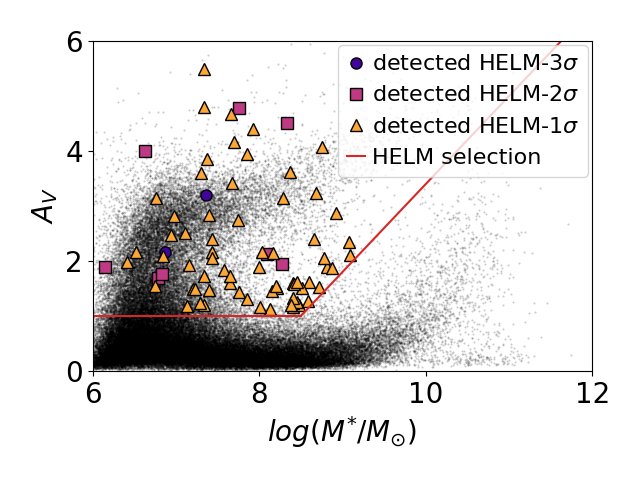}
    \caption{Stellar mass and dust extinction of all sources in the CEERS survey (black dots), as taken from Bisigello et al. (in prep.). The red line show the HELM selection criteria, while the coloured symbols shown the subsample of HELM sources that we expect to be detectable by PRIMAger in at least one filter. Different symbols indicate different HELM subsamples, namely HELM-$3\sigma$ (circles), HELM-$2\sigma$ (squares), and HELM-$1\sigma$ (triangles).  }
    \label{fig:MAV}
\end{figure}

The input HELM sample is taken from Bisigello et al. (in prep.). Briefly, the authors considered the latest data release (DR 0.7) of the public CEERS photometric catalogue \cite{Finkelstein2024} in the Extended Groth Strip (EGS) covering around $90\,\rm arcmin^{2}$. The catalogue includes observations in six HST filters (ACS/WFC F606W, F814W and WFC3/IR F105W, F125W, F140W, F160W), six broad-band JWST/NIRCam filters (F115W, F150W, F200W, F277W, F356W, F444W) and one JWST/NIRCam medium-band (F410M). The photometric data were fitted with the Bayesian Analysis of Galaxies for Physical Inference and Parameter EStimation (\textsc{Bagpipes} \cite{Carnall2018}) code using different star-formation histories (i.e. delayed, declining, and double-power-law), two different dust extinction laws (Calzetti et al. 2000 and Small Magellanic
Cloud)\cite{Calzetti2000,Gordon2003}, and including nebular continuum and emission lines using a range a ionisation parameter (i.e. dimensionless ratio of densities of ionising photons to hydrogen) from ${\rm log_{10}}(U)=-4$ to $-2$. Then, the HELM selection was performed considering the multiparameter probability distribution of $M_{*}$ and $A_V$ obtained from the SED fitting. In particular, the sample includes objects with the probability $68-95\%$ (HELM-$1\sigma$), $95-99.7\%$ (HELM-$2\sigma$), and $>99.7\%$ (HELM-$3\sigma$) of being either a galaxy with small stellar mass and $A_{V}>1$, or a galaxy with a larger mass and dust extinction above expectations. This selection, shown in Figure \ref{fig:MAV}, is described as:
\begin{align}\label{eq:selection}
    &({\rm log_{10}}(M_{*}/{\rm M_{\odot}})\leq8.5) \land (A_{V}>1) \;\lor\\
    &({\rm log_{10}}(M_{*}/{\rm M_{\odot}})>8.5) \land (A_{V}>1.6\,{\rm log_{10}}(M_{*}/{\rm M_{\odot}})-12.6).\nonumber
\end{align}
The sample was then cleaned by any source for which the best SED model with $A_V<1$ results on a $\chi^2$ lower than the best SED with larger dust extinction. The resulting catalogues were finally cross-matched with a public list of candidate brown-dwarfs \cite{Holwerda2024}, removing three sources, and $z>8$ galaxy candidates \cite{Finkelstein2024}, removing other four sources. The final catalogues include 2621 HELM-$1\sigma$, 174 HELM-$2\sigma$, and 17 HELM-$3\sigma$ objects. 

The HELM-$3\sigma$ sample includes galaxies up to $z=0.6$, with a median redshift of $z=0.3$. The HELM-$2\sigma$ sample contains mainly sources at $z\leq1$, with one galaxies with photometric redshift $z=2.9$ and one with spectroscopic redshift $z=4.8$ \cite{Bisigello2024}, but the median redshift of the HELM-$2\sigma$ sample is the same as that of the HELM-$3\sigma$ sample. HELM-$1\sigma$ sample includes the less robust candidates, their photometric redshifts extend up to $z=7.3$ and have a median value of $z=0.7$. 

%%%%%%%%%%%%%%%%%%%%%%%%%%%%%%%%%%%%%
\section{Predictions for PRIMA}\label{sec:predictions}

\begin{table}[]
    \caption{5$\sigma$ depth considered for each filter in PRIMAger. This is consistent with an integration time of 1000 hours over $1\,\rm deg^{2}$.}
    \centering
    \begin{tabular}{c|cccccc}
        filter & PHI1 & PHI2 & PPI1 & PPI2 & PPI3 & PPI4 \\
        \hline
       $\lambda [\mu m]$  &  24-45 & 45-84 & 96 & 126 & 172 & 235 \\
       5$\sigma$ depth [$\mu$Jy]  & 92-156 & 172-292 & 110 & 197 & 144 & 229 \\
    \end{tabular}

    \label{tab:depths}
\end{table}
We considered, for each HELM galaxy, the best SED-fitting model obtained from \textsc{Bagpipes} considering a delayed SFH, the Calzetti reddening law\cite{Calzetti2000}, the Draine \& Li (2007)\cite{Draine2007} dust emission model, an ionisation parameter between $\rm log_{10}(U)=-4$ and $-2$ and a Polycyclic Aromatic Hydrocarbon (PAH) mass fraction of 2. A previous study\cite{Bisigello2023b} have shown that changing the PAH mass fraction on a similar sample has a minor impact on the physical properties, with absolute differences in stellar mass and $A_{V}$ of 0.25 dex and 0.12 mag, respectively. 

From the best models, we extracted the fluxes expected in the PRIMAger filters. We considered the depths reached by integrating 1000\,h over $1\,\rm deg^{2}$, which are listed in Table \ref{tab:depths}. Given that at least 75\% of PRIMA 5-years time will be devoted to General Observer experiments, the chosen integration time corresponds to a feasible PRIMAger survey\cite{PRIMAGO}. However, it is necessary to mention that the $5\sigma$ confusion limits increase from $21\,\rm \mu Jy$ at $25\,\rm \mu m$ to $46\,\rm mJy$ at $235\,\rm \mu m$\cite{Bethermin2024}, exceeding the depths considered at $\lambda>42.6\,\rm \mu m$. Deblending techniques\cite{Donnellan2024} (such as \texttt{XID+}) are able to reduce confusion noise and still extract accurate fluxes for sources $\sim1\,\rm dex$ below the confusion limits at these wavelengths, provided sufficient prior information is available. In particular, it is possible to reach $5\sigma$ depths of $281\,\rm \mu Jy$ in $96\,\rm \mu m$ (PPI1), $747\,\rm \mu Jy$ at $126\,\rm \mu m$ (PPI2), $2650\,\rm \mu Jy$ in $172\,\rm \mu m$ (PPI3) and $7030\,\rm \mu Jy$ in $235\,\mu m$ (PPI4)\cite{Donnellan2024}. Moreover, further reductions in the confusion noise may also be possible using other techniques, such as applying an auto-encoder to super-resolve the maps (see, e.g., Koopmans et al. in prep) or starting from observations at higher angular resolutions, available at other wavelengths. Therefore, it will be fundamental to target areas already observed by optical sources, even if careful analysis may be necessary to cross-match data sets with different angular resolution using, for example, the likelihood ratio technique widely applied to previous FIR or radio data \cite{deRuiter1977,Sutherland1992,Ciliegi2003,Smith2011,McAlpine2012,Fleuren2012}. The same optical multi-wavelength data will also be fundamental to select HELM sources.

\begin{figure}
    \centering
    \includegraphics[width=0.75\linewidth]{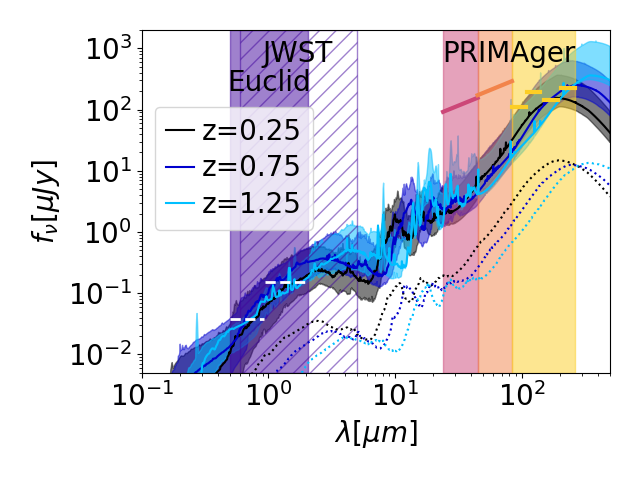}
    \caption{Median SED of the HELM sources that are expected to be detected in at least one PRIMAger filters in three redshift bins (solid lines). The shaded areas show the 1$\sigma$ dispersion of the SEDs, while the dotted lines show the median SED of the undetected sources in the same redshift bin. The vertical coloured bands show the wavelength coverage of \textit{JWST} (purple dashed area), \textit{Euclid} (purple area) and PRIMAger, divided between PHI1 (red area) , PHI2 (orange area) and PPI filters (yellow area). Horizontal solid lines show the survey depths reported in Table \ref{tab:depths}. The horizontal dashed lines show the $5\sigma$ depth expected over the $52\rm\deg^{2}$ of the Euclid Deep Survey.}
    \label{fig:sed}
\end{figure}

Considering the possible improvement in extracting sources below the classical confusion noise, we considered the $5\sigma$ instrumental depths and we looked for HELM galaxies with a $S/N>3$ in at least one PRIMAger filter. We obtained that 2/17 (12\%) HELM$-3\sigma$, 8/174 ($4\%$) HELM$-2\sigma$, and 67/2623 ($2\%$) HELM$-1\sigma$ galaxies are expected to be bright enough in the IR to be detected in at least one PRIMAger filter. The median SED of the HELM sources that we expect to be detectable by PRIMAger at redshift z=0.25, z=0.75, and z=1.25 is reported in Fig. \ref{fig:sed}. As visible in the same Figure, while \textit{JWST} is essential to trace the stellar continuum, PRIMAger will be able to trace the dust emission. In the same Figure we also show the wavelength range observed by \textit{Euclid} and the $5\sigma$ expected depths in the Euclid Deep Survey, which will cover $52\deg^{2}$ \cite{EuclidSkyOverview}. HELM sources will be detectable by \textit{Euclid} as well, allowing for deriving the stellar mass and the dust extinction, but also helping with the deblending of PRIMAger sources. The positions of the \textit{Euclid} fields is also ideal to have low FIR foreground contamination, since they avoid the galactic plane \cite{EuclidSkyOverview}.  

\begin{figure}
    \centering
    \includegraphics[width=0.57\linewidth]{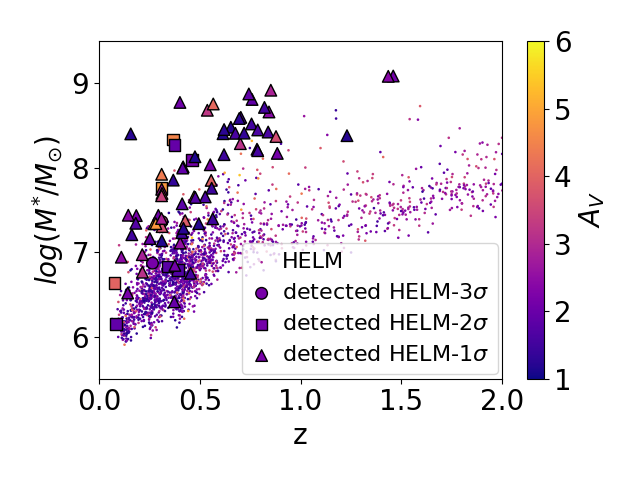}
    \includegraphics[width=0.40\linewidth]{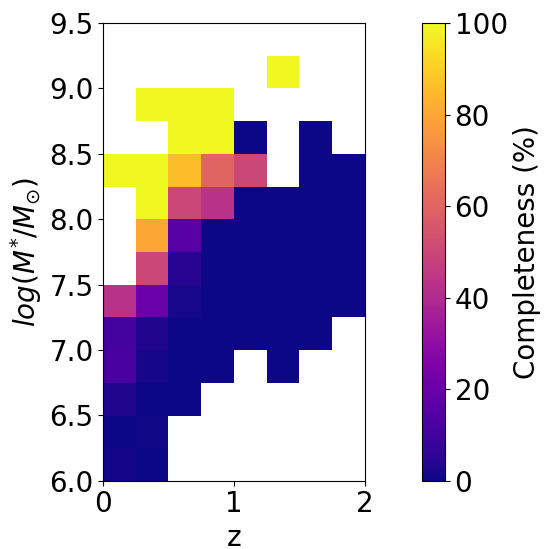}
    \caption{Left: Stellar mass and redshift of all HELM sources (coloured dots) and the subsamples expected to be detectable in at least one PRIMAger filter (coloured circles, squares, and triangles, as in the left panel). Symbols are colour-coded by their dust extinction. Right: completeness of HELM sources detected by PRIMAger as a function of redshift and stellar mass. }
    \label{fig:MzAV}
\end{figure}

As shown in Fig. \ref{fig:MAV} and \ref{fig:MzAV}, HELM galaxies that are expected to be detected by PRIMAger will be galaxies slightly more massive and less dust extincted with respect to the overall sample. In particular, the average stellar mass of the subsample detected by PRIMAger is expected to be ${\rm log_{10}}(M^{*}/\rm M_{\odot})=7.7$, while it is ${\rm log_{10}}(M^{*}/\rm M_{\odot})=7.1$ for the overall HELM sample. The average dust extinction is instead $A_{V}=1.9$ for detected HELM galaxies and $A_{V}=2.5$ for the overall sample. The average redshift of the detected HELM galaxies amounts to $z=0.4$, instead of $z=0.6$ considering the entire HELM sample.

\begin{figure}
    \centering
    \includegraphics[width=0.49\linewidth]{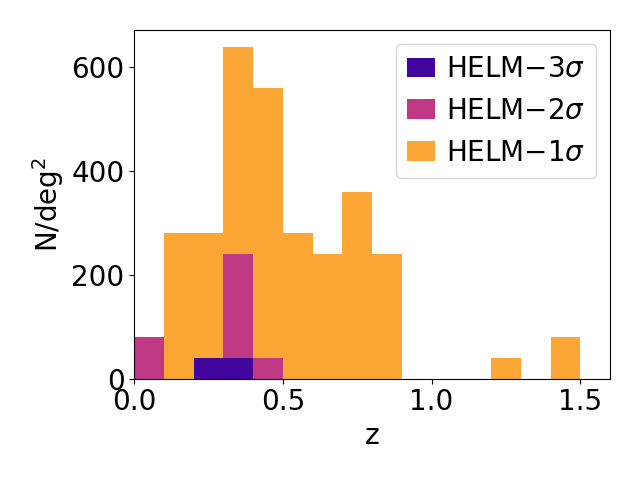}
    \includegraphics[width=0.49\linewidth]{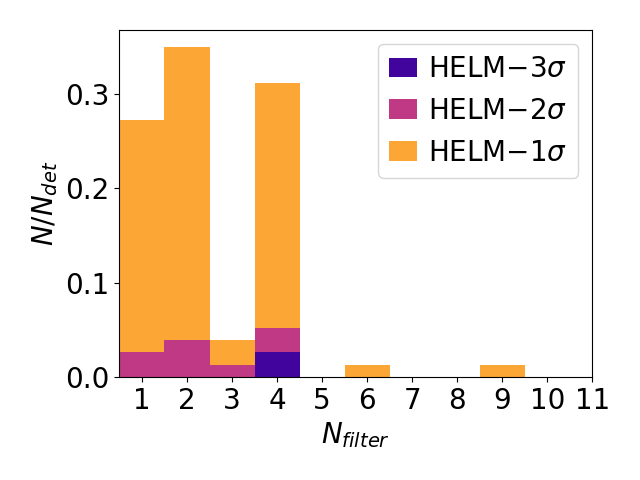}
    \caption{Left: Redshift distribution of HELM sources that are expected to be detected by at least one PRIMAger filter. We show the stack distribution of the three HELM subsample. Right: number of PRIMAger filters with a $S/N>3$ for each detected HELM source. }
    \label{fig:Nz}
\end{figure}

The fraction of HELM sources detected with PRIMAger is small, however PRIMAger will compensate by efficiently covering large sky areas. We can plan to observe fields already covered by optical and near-IR spectro-photometric data, like the one that will be part of the surveys by \textit{Euclid} \cite{Mellier2024} or the {\em Roman} Space Telescope. Optical and near-IR observations will be complementary to FIR ones, as they will allow for an estimation of the redshift and stellar mass of each source and a selection of HELM candidates. Overall, we expect PRIMAger to detect about $3.1\times10^{4}$ HELM sources over an area of 1$\,\rm deg^{2}$ (Fig. \ref{fig:Nz} left). The increase in number with respect to HELM sources already observed in CEERS with JWST is due to the increase in area, from $90\,\rm arcmin^2$ of CEERS to 1$\,\rm deg^{2}$. Sources observed with PRIMAger are expected to be mainly located at $z<1$, but we expect around 100 sources at $z=1-1.5$. In particular, around 70\% of HELMs in the detected sample will be observed in more than one PRIMAger filter, with 32\% of them being detected in the four longest wavelength PRIMAger filters (PPI1, PPI2, PPI3, PPI4; covering from $\lambda=90$ to $240\,\rm \mu m$). This wavelength coverage, combined with ancillary optical data like the one available with \textit{Euclid}, will allow to go beyond a simple FIR detection and perform a multi-wavelength SED modelling from optical to the FIR emission to estimate the SFR, the stellar mass, the dust extinction, and the dust mass. This will be fundamental to have a more comprehensive view of HELM galaxies and to understand if the large dust extinction observed in HELM galaxies is a geometry effect or is indeed due to a large dust content.

\section{Summary}\label{sec:summary}
In this paper, we considered a sample of HELM galaxies within the CEERS survey (Bisigello et al. in prep.), and investigated the capability of PRIMAger to observe them. This is crucial to investigate the dust content and properties of these peculiar objects, with the potential of shedding new light on dust production mechanisms of these galaxies. Moreover, constraining the properties of HELM galaxies can be helpful to filter out these objects from $z>15$ galaxies sample, since they are known lower-$z$ contaminants \cite{Gandolfi2025}. The HELM selection exploited here was made considering the multi-parameter distribution of stellar mass and dust extinction, derived by fitting the available HST and JWST photometric data with several setups (Bisigello et al., in prep.). The SED fitting models were then used to derive the fluxes expected in the FIR and to compare them with the depths expected in a PRIMAger survey of 1000 hours over $1\,\rm deg^{2}$, which correspond to a feasible size and depth for a PRIMAger survey. 

PRIMAger will be able to observe the massive tail of HELM galaxies, with an average stellar mass of ${\rm log_{10}}(M^{*}/\rm M_{\odot})=7.7$ and an average dust extinction of $A_{V}=1.9$. These galaxies represent a minority of the HELM population, around 3\% of the total HELM sample, but PRIMAger can leverage this small number with the area coverage. In particular, considering a deep field of 1000h over $1\,\deg^2$, we expect to observe $3.1 \times 10^4$ HELM galaxies, of which 100 at $z=1-1.5$. Stacking analysis can be further considered to derive estimates of the FIR fluxes for fainter sources, but stacking also requires observations at optical wavelengths to pre-select HELM sources. For this reason, observing fields with public optical images, like the Euclid Deep Fields, will be fundamental for selecting HELM galaxies, derive their stellar mass and $A_{V}$, and help deblending PRIMAger photometry. 

Moreover, the power of PRIMAger will not be limited to a simple detection of these sources in the FIR, but it will allow for characterising their obscured SFR and dust properties. In fact, 32\% of the detected HELM sources will have more than four detections in PRIMAger (mainly at $\lambda=90-240\,\rm \mu m$), allowing for a full characterization of their dust emission. Therefore, the combination of optical and FIR data of HELM galaxies will be crucial for a detailed characterization of their ISM and their physical properties, comparing, for example, their dust mass to their metal content. 

\subsection*{Disclosures} 
The authors have no relevant financial interests in the manuscript or other potential conflicts of interest.

\subsection*{Code, Data, and Materials Availability}
The \textsc{Bagpipes} code is publicly available on this \href{https://bagpipes.readthedocs.io/en/latest/}{link}. The CEERS data used for the predictions are also publicly available through the \href{https://ceers.github.io/}{CEERS website}.

\subsection* {Acknowledgments}
The research activities described in this paper were carried out with contribution of the Next Generation EU funds within the National Recovery and Resilience Plan (PNRR), Mission 4 - Education and Research, Component 2 - From Research to Business (M4C2), Investment Line 3.1 - Strengthening and creation of Research Infrastructures, Project IR0000034 – “STILES - Strengthening the Italian Leadership in ELT and SKA”

%%%%% References %%%%%

\bibliography{report}   % bibliography data in report.bib

\begin{thebibliography}{10}

\bibitem{Draine2003}
B.~T. {Draine}, ``{Interstellar Dust Grains},'' {\em \araa} {\bf 41}, 241--289  (2003).

\bibitem{Hauser2001}
M.~G. {Hauser} and E.~{Dwek}, ``{The Cosmic Infrared Background: Measurements and Implications},'' {\em \araa} {\bf 39}, 249--307  (2001).

\bibitem{Madau2014}
P.~{Madau} and M.~{Dickinson}, ``{Cosmic Star-Formation History},'' {\em \araa} {\bf 52}, 415--486  (2014).

\bibitem{Gruppioni2020}
C.~{Gruppioni}, M.~{B{\'e}thermin}, F.~{Loiacono}, {\em et~al.}, ``{The ALPINE-ALMA [CII] survey. The nature, luminosity function, and star formation history of dusty galaxies up to z $\sim$ 6},'' {\em \aap} {\bf 643}, A8  (2020).

\bibitem{Zavala2021}
J.~A. {Zavala}, C.~M. {Casey}, S.~M. {Manning}, {\em et~al.}, ``{The Evolution of the IR Luminosity Function and Dust-obscured Star Formation over the Past 13 Billion Years},'' {\em \apj} {\bf 909}, 165  (2021).

\bibitem{Traina2024}
A.~{Traina}, C.~{Gruppioni}, I.~{Delvecchio}, {\em et~al.}, ``{A$^{3}$COSMOS: The infrared luminosity function and dust-obscured star formation rate density at 0.5 < z < 6},'' {\em \aap} {\bf 681}, A118  (2024).

\bibitem{Wang2019}
T.~{Wang}, C.~{Schreiber}, D.~{Elbaz}, {\em et~al.}, ``{A dominant population of optically invisible massive galaxies in the early Universe},'' {\em \nat} {\bf 572}, 211--214  (2019).

\bibitem{Gruppioni2010}
C.~{Gruppioni}, F.~{Pozzi}, P.~{Andreani}, {\em et~al.}, ``{PEP: First Herschel probe of dusty galaxy evolution up to z \raisebox{-0.5ex}\textasciitilde 3},'' {\em \aap} {\bf 518}, L27  (2010).

\bibitem{Talia2021}
M.~{Talia}, A.~{Cimatti}, M.~{Giulietti}, {\em et~al.}, ``{Illuminating the Dark Side of Cosmic Star Formation Two Billion Years after the Big Bang},'' {\em \apj} {\bf 909}, 23  (2021).

\bibitem{Gentile2024}
F.~{Gentile}, M.~{Talia}, M.~{Behiri}, {\em et~al.}, ``{Illuminating the Dark Side of Cosmic Star Formation. III. Building the Largest Homogeneous Sample of Radio-selected Dusty Star-forming Galaxies in COSMOS with PhoEBO},'' {\em \apj} {\bf 962}, 26  (2024).

\bibitem{Pannella2015}
M.~{Pannella}, D.~{Elbaz}, E.~{Daddi}, {\em et~al.}, ``{GOODS-Herschel: Star Formation, Dust Attenuation, and the FIR-radio Correlation on the Main Sequence of Star-forming Galaxies up to z $\sim$4},'' {\em \apj} {\bf 807}, 141  (2015).

\bibitem{McLure2018}
R.~J. {McLure}, J.~S. {Dunlop}, F.~{Cullen}, {\em et~al.}, ``{Dust attenuation in 2 < z < 3 star-forming galaxies from deep ALMA observations of the Hubble Ultra Deep Field},'' {\em \mnras} {\bf 476}, 3991--4006  (2018).

\bibitem{Shapley2023}
A.~E. {Shapley}, R.~L. {Sanders}, N.~A. {Reddy}, {\em et~al.}, ``{JWST/NIRSpec Balmer-line Measurements of Star Formation and Dust Attenuation at z 3-6},'' {\em \apj} {\bf 954}, 157  (2023).

\bibitem{Liu2024}
Z.~{Liu}, T.~{Morishita}, and T.~{Kodama}, ``{Characterizing Dust Extinction and Spatially Resolved Paschen-$\alpha$ Emission within 97 Galaxies at $1<z<1.6$ with JWST NIRCam Slitless Spectroscopy},'' {\em arXiv e-prints} , arXiv:2406.11188  (2024).

\bibitem{Rodighiero2023}
G.~{Rodighiero}, L.~{Bisigello}, E.~{Iani}, {\em et~al.}, ``{JWST unveils heavily obscured (active and passive) sources up to z 13},'' {\em \mnras} {\bf 518}, L19--L24  (2023).

\bibitem{Bisigello2023b}
L.~{Bisigello}, G.~{Gandolfi}, A.~{Grazian}, {\em et~al.}, ``{Delving deep: A population of extremely dusty dwarfs observed by JWST},'' {\em \aap} {\bf 676}, A76  (2023).

\bibitem{Bisigello2024}
L.~{Bisigello}, G.~{Gandolfi}, A.~{Feltre}, {\em et~al.}, ``{Spectroscopic confirmation of a dust-obscured, metal-rich dwarf galaxy at z\raisebox{-0.5ex}\textasciitilde5},'' {\em arXiv e-prints} , arXiv:2410.10954  (2024).

\bibitem{Gandolfi2025}
G.~{Gandolfi}, G.~{Rodighiero}, L.~{Bisigello}, {\em et~al.}, ``{Ultra High-Redshift or Closer-by, Dust-Obscured Galaxies? Deciphering the Nature of Faint, Previously Missed F200W-Dropouts in CEERS},'' {\em arXiv e-prints} , arXiv:2502.02637  (2025).

\bibitem{Gall2011}
C.~{Gall}, J.~{Hjorth}, and A.~C. {Andersen}, ``{Production of dust by massive stars at high redshift},'' {\em \aapr} {\bf 19}, 43  (2011).

\bibitem{Sarangi2018}
A.~{Sarangi}, M.~{Matsuura}, and E.~R. {Micelotta}, ``{Dust in Supernovae and Supernova Remnants I: Formation Scenarios},'' {\em \ssr} {\bf 214}, 63  (2018).

\bibitem{Dayal2013}
P.~{Dayal}, A.~{Ferrara}, and J.~S. {Dunlop}, ``{The physics of the fundamental metallicity relation},'' {\em \mnras} {\bf 430}, 2891--2895  (2013).

\bibitem{Bagley2023}
M.~B. {Bagley}, S.~L. {Finkelstein}, A.~M. {Koekemoer}, {\em et~al.}, ``{CEERS Epoch 1 NIRCam Imaging: Reduction Methods and Simulations Enabling Early JWST Science Results},'' {\em \apjl} {\bf 946}, L12  (2023).

\bibitem{Yang2023}
G.~{Yang}, C.~{Papovich}, M.~B. {Bagley}, {\em et~al.}, ``{CEERS MIRI Imaging: Data Reduction and Quality Assessment},'' {\em \apjl} {\bf 956}, L12  (2023).

\bibitem{Finkelstein2024}
S.~L. {Finkelstein}, G.~C.~K. {Leung}, M.~B. {Bagley}, {\em et~al.}, ``{The Complete CEERS Early Universe Galaxy Sample: A Surprisingly Slow Evolution of the Space Density of Bright Galaxies at z {\ensuremath{\sim}} 8.5{\textendash}14.5},'' {\em \apjl} {\bf 969}, L2  (2024).

\bibitem{Chabrier2003}
G.~{Chabrier}, ``{Galactic Stellar and Substellar Initial Mass Function},'' {\em \pasp} {\bf 115}, 763--795  (2003).

\bibitem{Carnall2018}
A.~C. Carnall, R.~J. McLure, J.~S. Dunlop, {\em et~al.}, ``{Inferring the star formation histories of massive quiescent galaxies with bagpipes: evidence for multiple quenching mechanisms},'' {\em MNRAS} {\bf 480}, 4379--4401  (2018).

\bibitem{Calzetti2000}
D.~{Calzetti}, L.~{Armus}, R.~C. {Bohlin}, {\em et~al.}, ``{The Dust Content and Opacity of Actively Star-forming Galaxies},'' {\em \apj} {\bf 533}, 682--695  (2000).

\bibitem{Gordon2003}
K.~D. {Gordon}, G.~C. {Clayton}, K.~A. {Misselt}, {\em et~al.}, ``{A Quantitative Comparison of the Small Magellanic Cloud, Large Magellanic Cloud, and Milky Way Ultraviolet to Near-Infrared Extinction Curves},'' {\em \apj} {\bf 594}, 279--293  (2003).

\bibitem{Holwerda2024}
B.~W. {Holwerda}, C.-C. {Hsu}, N.~{Hathi}, {\em et~al.}, ``{Cosmic evolution early release science survey (CEERS): multiclassing galactic dwarf stars in the deep JWST/NIRCam},'' {\em \mnras} {\bf 529}, 1067--1081  (2024).

\bibitem{Draine2007}
B.~T. {Draine} and A.~{Li}, ``{Infrared Emission from Interstellar Dust. IV. The Silicate-Graphite-PAH Model in the Post-Spitzer Era},'' {\em \apj} {\bf 657}, 810--837  (2007).

\bibitem{PRIMAGO}
A.~{Moullet}, T.~{Kataria}, D.~{Lis}, {\em et~al.}, ``{PRIMA General Observer Science Book},'' {\em arXiv e-prints} , arXiv:2310.20572  (2023).

\bibitem{Bethermin2024}
M.~{B{\'e}thermin}, A.~D. {Bolatto}, F.~{Boulanger}, {\em et~al.}, ``{Confusion of extragalactic sources in the far-infrared: A baseline assessment of the performance of PRIMAger in intensity and polarization},'' {\em \aap} {\bf 692}, A52  (2024).

\bibitem{Donnellan2024}
J.~M.~S. Donnellan, S.~J. Oliver, M.~Béthermin, {\em et~al.}, ``Overcoming confusion noise with hyperspectral imaging from primager,'' {\em Monthly Notices of the Royal Astronomical Society} {\bf 532}, 1966--1979  (2024).

\bibitem{deRuiter1977}
H.~R. {de Ruiter}, A.~G. {Willis}, and H.~C. {Arp}, ``{A Westerbork 1415 MHz survey of background radio sources. II. Optical identifications with deep IIIa-J plates.},'' {\em \aaps} {\bf 28}, 211--293  (1977).

\bibitem{Sutherland1992}
W.~{Sutherland} and W.~{Saunders}, ``{On the likelihood ratio for source identification.},'' {\em \mnras} {\bf 259}, 413--420  (1992).

\bibitem{Ciliegi2003}
P.~{Ciliegi}, G.~{Zamorani}, G.~{Hasinger}, {\em et~al.}, ``{A deep VLA survey at 6 cm in the Lockman Hole},'' {\em \aap} {\bf 398}, 901--918  (2003).

\bibitem{Smith2011}
D.~J.~B. {Smith}, L.~{Dunne}, S.~J. {Maddox}, {\em et~al.}, ``{Herschel-ATLAS: counterparts from the ultraviolet-near-infrared in the science demonstration phase catalogue},'' {\em \mnras} {\bf 416}, 857--872  (2011).

\bibitem{McAlpine2012}
K.~{McAlpine}, D.~J.~B. {Smith}, M.~J. {Jarvis}, {\em et~al.}, ``{The likelihood ratio as a tool for radio continuum surveys with Square Kilometre Array precursor telescopes},'' {\em \mnras} {\bf 423}, 132--140  (2012).

\bibitem{Fleuren2012}
S.~{Fleuren}, W.~{Sutherland}, L.~{Dunne}, {\em et~al.}, ``{Herschel-ATLAS: VISTA VIKING near-infrared counterparts in the Phase 1 GAMA 9-h data},'' {\em \mnras} {\bf 423}, 2407--2424  (2012).

\bibitem{EuclidSkyOverview}
Y.~{Euclid Collaboration: Mellier}, {Abdurro'uf}, J.~{Acevedo~Barroso}, {\em et~al.}, ``{Euclid. I. Overview of the Euclid mission},'' {\em \aap, accepted} , arXiv:2405.13491  (2024).

\bibitem{Mellier2024}
Y.~{Euclid Collaboration: Mellier}, {Abdurro'uf}, J.~A. {Acevedo Barroso}, {\em et~al.}, ``{Euclid. I. Overview of the Euclid mission},'' {\em arXiv e-prints} , arXiv:2405.13491  (2024).

\end{thebibliography}
\bibliographystyle{spiejour}   % makes bibtex use spiejour.bst

\end{spacing}
\end{document}